%% file: Data-Driven_SMC.tex
\documentclass[twocolumn]{autart} 

\usepackage{amsmath,amsfonts,amssymb,mathrsfs}
\usepackage{graphicx}
\usepackage{natbib}        
\usepackage{enumerate}
\usepackage{wrapfig}
\usepackage{subcaption}

\newtheorem{assume}{Assumption}[section]
\newtheorem{theorem}{Theorem}[section]

\newtheorem{lemma}{Lemma}[section]

\newtheorem{proof}{Proof}

\newtheorem{proposition}{Proposition}[section]

\begin{document}
	
	\begin{frontmatter}
		\title{Data-driven sliding mode control for partially unknown nonlinear systems}
		
		
		\author[1,cor]{Jianglin Lan}\ead{Jianglin.Lan@glasgow.ac.uk},
		\corauth[cor]{Corresponding author}
		\author[2]{Xianxian Zhao}\ead{xianxian.zhao@ucd.ie},	
		\author[3]{Congcong Sun}\ead{congcong.sun@wur.nl},
		\address[1]{James Watt School of Engineering, University of Glasgow, Glasgow G12 8QQ, United Kingdom}
		\address[2]{School of Electrical and Electronic Engineering, University College Dublin, Belfield, D04 V1W8 Dublin, Ireland}
		\address[3]{Agricultural Biosystems Engineering Group, Wageningen University and Research, 6700 AC Wageningen, The Netherlands}
		

		\begin{abstract}
			\input{sections/abstract}
		\end{abstract}
	\end{frontmatter}

\section{Introduction}

\input{sections/introduction}

\section{Problem description} \label{sec:problem statement}
\input{sections/preliminary}

\section{Data-driven sliding mode control}\label{sec:main results}
\input{sections/overview}

\subsection{Reachability and convergence of sliding surface}\label{sec:SMC controller}
\input{sections/SMC_control}

\subsection{Data-driven nominal controller design}\label{sec:nominal control}
\input{sections/nominal_control}

\section{Simulation results} \label{sec:simulation}
\input{sections/simulation}

\section{Conclusion} \label{sec:conclusion}
\input{sections/conclusion}

\bibliographystyle{model5-names}
\bibliography{References}

\end{document}

%% file: sections/abstract.tex
This paper presents a new data-driven control for multi-input, multi-output nonlinear systems with partially unknown dynamics and bounded disturbances. Since exact nonlinearity cancellation is not feasible with unknown disturbances, we adapt sliding mode control (SMC) for system stability and robustness. The SMC features a data-driven robust controller to reach the sliding surface and a data-driven nominal controller from a semidefinite program (SDP) to ensure stability. Simulations show the proposed method outperforms existing data-driven approaches with approximate nonlinearity cancellation.

%% file: sections/introduction.tex
Model-based control struggles with systems too complex or uncertain for precise modelling, as it relies on accurate system identification. Data-driven methods overcome this by designing controllers directly from plant data \citep{TangDaoutidis22,berberich2024overview}, enabling more adaptive strategies in fields like biology, soft robotics, and industry. 

Notable data-driven control methods include adaptive control \citep{Astolfi21n}, virtual reference feedback tuning \citep{CampiSavaresi06}, adaptive dynamic programming \citep{LewisVrabie09}, and the system behaviour approach \citep{PersisTesi20}. 
Despite this, designing data-driven control for nonlinear systems remains challenging, particularly in ensuring theoretical guarantees and computational feasibility with finite data \citep{PersisTesi23}. 

Existing data-driven approaches for nonlinear systems use behavioural theory, set membership, kernel methods, Koopman operator, or feedback linearization \citep{Martin+23}. A key approach from \cite{PersisTesi20} represents plant dynamics via system trajectories and solves a data-dependent semidefinite program (SDP) for controller synthesis. This method designs state-feedback control to stabilize the system around equilibrium using a Taylor approximation, assuming a linearly bounded remainder. Later works \cite{Martin+23} extend it by incorporating the remainder for global stabilization, but assuming disturbance-free systems. Polynomial approximation is also used for continuous-time nonlinear systems \citep{Guo+22d, Martin+23d}, but with faster vanishing remainders ensuring only local stabilization. Recent works \citep{Persis+23, Guo+23} use data-driven control with approximate nonlinearity cancellation (referred to as \textit{data-ANC}) for local stabilization, implicitly mitigating disturbance effects via regularization.

This paper enhances the robustness performance of the \textit{data-ANC} method \citep{Persis+23, Guo+23} by proposing a novel approach to globally stabilize nonlinear systems with partially unknown dynamics and disturbances.
The main contributions are as follows:
\begin{itemize}
	\item We propose a data-driven sliding mode control (SMC) for global stabilization of nonlinear systems, extending beyond the local stability of approximation-based methods \citep{PersisTesi20, Guo+22d, Martin+23d}. Unlike prior model-free SMC approaches limited to single-input single-output systems \citep{ebrahimi+20, Corradini21, riva2024data}, our method handles multi-input multi-output systems.
	\item The proposed SMC uses a nominal controller whose gain is solved from a data-dependent SDP based on $H_\infty$ robust control. Unlike the \textit{data-ANC} method \citep{Persis+23, Guo+23}, our method formally ensures robustness against disturbances, rather than mitigating their effects through regularization. Empirical studies also demonstrate improved SDP feasibility.
\end{itemize}

The rest of this paper is structured as follows:
Section \ref{sec:problem statement} describes the control problem, Section \ref{sec:main results} presents the proposed data-driven SMC, Section \ref{sec:simulation} 
reports the simulation results, and Section \ref{sec:conclusion} draws the 
conclusions.

%% file: sections/preliminary.tex
Consider the discrete-time nonlinear control system
\begin{equation} \label{eq:sys origin}
	x(k+1) = A_x x(k) + A_q Q(x(k)) + B u(k) + D w(k),
\end{equation}
where $x(k) \in \mathbb{R}^{n_x}$ is the state, $u(k) \in \mathbb{R}^{n_u}$ is the control input, and $w(k)  \in \mathbb{R}^{n_w}$ is the disturbance. $Q(x(k)) \in \mathbb{R}^{n_q}$ contains only the nonlinear functions of $x(k)$. $A_x$ and $A_q$ are unknown constant matrices, while $B$ and $D$ are assumed to be known.
The disturbance $w(k)$ is unknown but bounded as in Assumption \ref{assume:disturbance}.
\begin{assume}\label{assume:disturbance}
	$|w| \leq \delta \times \mathbf{1}_{n_w}$ for some known $\delta > 0$.	
\end{assume} 

By defining $Z(x(k)) = [x(k)^\top, ~Q(x(k))^\top]^\top$, system \eqref{eq:sys origin} can be compactly represented as
\begin{equation} \label{eq:sys for design}
	x(k+1) = A Z(x(k)) + B u(k) + D w(k),
\end{equation}
with the unknown matrix $A = [A_x, A_q]$.

Given the presence of nonlinearity, disturbance, and unknown matrix $A$, this paper designs a data-driven controller to robustly stabilize system \eqref{eq:sys for design}, or equivalently \eqref{eq:sys origin}, using
sliding mode control and $H_\infty$ control theories based on
collected data sequences of $x(k)$ and $u(k)$.

%% file: sections/overview.tex
The controller $u(k)$ is designed as
\begin{equation}\label{eq:overall controller}
	u(k) = u_\text{n}(k) + u_\text{r}(k)
\end{equation}
with a nominal controller $u_\text{n}(k)$ and a robust controller $u_\text{r}(k)$ in the forms of
\begin{align}\label{eq:ur}
	u_\text{n}(k) ={}& K Z(x(k)), \nonumber\\
	u_\text{r}(k)  ={}& (N B)^{\dagger} [- \tilde{A} Z(x(k)) + (1-q) \phi(k) s(k) \nonumber\\
	&- \varphi(k) \cdot \mathrm{sgn}(s(k)) ], 
\end{align}
where the gain $K$ and constant matrix $\tilde{A}$ are detailed in Section \ref{sec:nominal control} (see Theorem~\ref{thm:control design} and Proposition \ref{proposition:condition}). The sliding variable, $s(k) \in \mathbb{R}^{m}$, is designed as 
\begin{equation}\label{eq:sliding surface}
	s(k) = N x(k),
\end{equation}
where $N \in \mathbb{R}^{m \times n_x}$, $m \leq n_u$, is chosen such that $N B$ is of full row rank $m$ with the pseudo-inverse $(N B)^\dagger$. $\mathrm{sgn}(\cdot)$ is the signum function.
The scalar $q$ is chosen such that $0 < q < 1$. The $m \times m$ diagonal matrices $\phi(k)$ and $\varphi(k)$ have diagonal entries, $\phi_{i,i}(k)$ and $\varphi_{i,i}(k)$, respectively, designed as follows:
\begin{equation}\label{eq:paras}
	\phi_{i,i}(k) = \frac{2}{\mathrm{e}^{-\sigma s_i(k)} + \mathrm{e}^{\sigma s_i(k)}}, ~
	\varphi_{i,i}(k) = \rho_i |s_i(k)|, 
\end{equation}
with constants $\sigma > 0$ and $0 < \rho_i < 1$. 
It can be observed that $0 < \phi_{i,i}(k) \leq 1$.

%% file: sections/SMC_control.tex
This section shows that the controller in \eqref{eq:overall controller} drives the state to the sliding surface $s(k) = 0$ and keeps it there. 
We adopt the simple yet conservative reaching conditions from Lemma \ref{lemma:sliding reachable conds} to illustrate the key ideas, leaving more advanced alternatives \citep{lesniewski2018discrete} for interested readers.

\begin{lemma}\citep{Sarpturk+87} \label{lemma:sliding reachable conds}
For discrete-time SMC, the sliding surface $s(k) = 0$ is reachable and convergent if and only if
	\begin{subequations}\label{SMC conds}
		\begin{align}
			\label{SMC cond1}
			(s_i(k+1) - s_i(k)) \cdot \mathrm{sgn}(s_i(k)) &< 0, ~ i \in [1,m], \\
			\label{SMC cond2}
			(s_i(k+1) + s_i(k)) \cdot \mathrm{sgn}(s_i(k)) &> 0, ~ i \in [1,m].
		\end{align}
	\end{subequations} 
\end{lemma}
Combining the inequalities in \eqref{SMC conds} yields $|s(k+1)| < |s(k)|$, ensuring convergence to $s(k) = 0$.
Theorem \ref{theorem:sliding and convergence} confirms the proposed controller satisfies \eqref{SMC conds}. 

\begin{theorem}\label{theorem:sliding and convergence}
The states of system \eqref{eq:sys for design} reach and stay near the sliding surface $s(k) = 0$, within the set  
\begin{equation}\label{set:theorem sliding}
	\Omega = \{ s(k) \in \mathbb{R}^{m} \mid  |s_i(k)| \leq \lambda_i \bar{f}_i, i \in [1,m]\},
\end{equation}
where $\lambda_i = \max (1/(2 - q - \rho_i), 1/(q+\rho_i))$ and $\bar{f}_i$ bounds the $i$-th element of $N D (d(k) + w(k))$, i.e.
 $|f_i| \leq \bar{f}_i$, 
if the nominal controller $u_\text{n}(k)$ is designed such that
\begin{equation}\label{cond:lemma sliding}
\!\! N A Z(x(k)) + NB K Z(x(k)) \!=\! \tilde{A} x(k) + N D d(k),\!\!\! 
\end{equation}
where $\tilde{A}$ is a constant matrix related to the nominal control and $d(k)$ is a lumped disturbance related to $w(k)$ and $Z(x(k))$, as detailed in Section \ref{sec:nominal control}. 
\end{theorem}

\begin{proof}
From \eqref{eq:sys for design}, we have
\begin{equation}\label{eq:s dot}
	s(k+1) = N A Z(x(k)) + N B u(k) + N D w(k).
\end{equation}
Applying \eqref{cond:lemma sliding} to \eqref{eq:s dot} yields
\begin{align}\label{eq:s dot2}
	& s(k+1) \nonumber\\
	={}& \tilde{A} Z(x(k)) + N D d(k) - \tilde{A} Z(x(k))  \nonumber\\
	& + (1-q) \phi(k) s(k) - \varphi(k) \cdot \mathrm{sgn}(s(k)) + N D w(k) \nonumber\\
	={}& (1-q) \phi(k) s(k) - \varphi(k) \cdot \mathrm{sgn}(s(k)) + N D \tilde{d}(k), 
\end{align}
where $\tilde{d}(k) = d(k) + w(k)$.

To demonstrate that the sliding surface is reachable, we prove \eqref{SMC conds} in Lemma \ref{lemma:sliding reachable conds}.
From \eqref{eq:sliding surface} - \eqref{eq:paras} and \eqref{eq:s dot2}, for any $i \in [1,m]$, we derive
\begin{align}\label{eq:cond1 pf1}
	& (s_i(k+1) - s_i(k)) \cdot \mathrm{sgn}(s_i(k)) \nonumber\\
	={}& [(1-q) \phi_{i,i}(k) s_i(k) - \varphi_{i,i}(k) \cdot \mathrm{sgn}(s_i(k)) - s_i(k) + f_i ]  \nonumber\\
	&\cdot   \mathrm{sgn}(s_i(k))  \nonumber\\
	={}& [(1-q) \phi_{i,i}(k) - \rho_i - 1] |s_i(k)| + f_i \cdot \mathrm{sgn}(s_i(k)) \nonumber\\
	\leq{}& -(q + \rho_i) |s_i(k)| + \bar{f}_i,
\end{align}
where $f_i$ is the $i$-th element of $N D \tilde{d}(k)$, with $|f_i| \leq \bar{f}_i$. 
Since $0 < q < 1$ and $0 < \rho_i < 1$, the inequality \eqref{SMC cond1} holds when $|s_i(k)| > \bar{f}_i / (q + \rho_i)$. 

Similarly, we derive from \eqref{eq:sliding surface} - \eqref{eq:paras} and \eqref{eq:s dot2} that
\begin{align}\label{eq:cond2 pf1}
& (s_i(k+1) + s_i(k)) \cdot \mathrm{sgn}(s_i(k)) \nonumber\\
\geq{}& (1-q) \phi_{i,i}(k) |s_i(k)| + (1 - \rho_i) |s_i(k)| - \bar{f}_i \nonumber\\
\geq{}& (2 - q - \rho_i)|s_i(k)| - \bar{f}_i, 
\end{align}
where $0 < q < 1$ and $0 < \phi_{i,i}(k) \leq 1$ are used.
Since $0 < q + \rho_i < 2$, it follows from \eqref{eq:cond2 pf1} that the inequality \eqref{SMC cond2} holds for $|s_i(k)| > \bar{f}_i / (2 - q - \rho_i)$.

In summary, the inequalities in \eqref{SMC conds} hold when $|s_i(k)| > \lambda_i \bar{f}_i$, where $\lambda_i = \max (1/(2 - q - \rho_i), 1/(q+\rho_i))$. This ensures the system states reach and stay near the sliding surface $s(k) = 0$, within the set $\Omega$ in \eqref{set:theorem sliding}. \qed
\end{proof}

Since $\bar{f}_i$ depends on the user-chosen matrix $N$, the size of $\Omega$ can be tuned via $N$, making $\Omega$ arbitrarily small.
By the sliding dynamics \eqref{eq:s dot}, the equivalent robust controller $u_\text{r}^{\text{eq}}(k) = -(N B)^{\dagger} N D w(k)$ aims to cancel the disturbance. Thus, the overall equivalent controller is 
\begin{equation}\label{eq:equivalent control}
	u_\mathrm{eq} = u_\text{r}^{\text{eq}}(k) + u_\text{n}(k).
\end{equation}
Substituting \eqref{eq:equivalent control} into \eqref{eq:s dot} gives
\begin{equation}\label{eq:s dot equivalent}
	s(k+1) = N (A Z(x(k)) + B u_\text{n}(k)).
\end{equation}
Hence, convergence of the sliding dynamics is ensured by designing $u_\text{n}(k)$ such that $A Z(x(k)) + B u_\text{n}(k)$ is robustly asymptotically stable, as detailed in Section \ref{sec:nominal control}.

%% file: sections/nominal_control.tex
Substituting the equivalent control \eqref{eq:equivalent control} into \eqref{eq:sys for design} gives
\begin{equation}\label{eq:nominal closed sys}
	x(k+1) = A Z(x(k)) + B u_\text{n}(k) + \Phi D w(k),
\end{equation}
where $\Phi = I_{n_x} - B (N B)^{\dagger} N$.

The nominal controller $u_\text{n}(k) = K Z(x(k))$ should be designed to ensure robust stability of system \eqref{eq:nominal closed sys}, keeping $x(k)$ stable within the set $\Omega$. Since the matrix $A$ is unknown, the gain $K$ is computed using a data-driven method. 
To this end, we derive a data-based representation of this system using 
$T$ collected samples, which satisfy \eqref{eq:sys for design} as follows:
\begin{equation}\label{eq:data dynamics}
	x(t+1) = A Z(x(t)) + B u(t) + D w(t), ~t \in [0,T-1].	
\end{equation}
These samples are grouped into the data sequences:	
\begin{align}	
		U_0 &= [u(0), u(1), \cdots, u(T-1)] \in \mathbb{R}^{n_u \times T}, \nonumber\\
		X_0 &= [x(0), x(1), \cdots, x(T-1)] \in \mathbb{R}^{n_x \times T}, \nonumber\\
		X_1 &= [x(1), x(2), \cdots, x(T)] \in \mathbb{R}^{n_x \times T}, \nonumber\\
	\!Z_0 &= [Z(x(0)), Z(x(1)), \cdots, Z(x(T-1))] \!\in\! \mathbb{R}^{n_z \times T}. \nonumber
\end{align}	
The corresponding disturbance sequence is
$W_0 = [w(0),w(1),\cdots,w(T-1)] \in \mathbb{R}^{n_w \times T}
$, which is unknown but bounded under Assumption \ref{assume:disturbance} \citep[Lemma 4]{Persis+23}.
The proposed data-driven nominal control design is presented in
Theorem~\ref{thm:control design}.
\begin{theorem}\label{thm:control design}
	Under Assumption \ref{assume:disturbance}, system \eqref{eq:nominal closed sys} is robustly stable with $u_\text{n}(k) = K Z(x(k))$ and gain 
	\begin{equation}\label{controller:thm}
		K = U_0[Y, ~ G_2] \left(\text{diag}(P, I_{n_z-n_x}) \right)^{-1},
	\end{equation} 
	with the matrices $P \in \mathbb{R}^{n_x \times n_x}$, $Y \in \mathbb{R}^{T \times n_x}$, and $G_2 \in \mathbb{R}^{T \times (n_z - n_x)}$ obtained from the SDP problem:
	\begin{subequations}\label{op:control design}	
		\begin{align}
			&\hspace{1.8cm}  \underset{P, Y, G_2, \gamma}{\min}~\gamma \nonumber\\
			\label{const:control 1}
			&\text{subject~to:} ~ 
			P \succ 0, ~ \gamma > 0, \\
			\label{const:control 2}
			&\hspace{1.6cm} Z_0 [Y, G_2] = \text{diag}(P, I_{n_z-n_x})
			\\
			\label{const:control 4}
			&\hspace{1.6cm} (X_1 + B U_0) G_2 = \mathbf{0}, \\
			\label{const:control 5}
			&
			\setlength\arraycolsep{1pt}
			\begin{bmatrix}
				P  & \mathbf{0} & P & \Upsilon_{1,4} & \mathbf{0} & Y^\top & \mathbf{0}\\
				\star & \gamma I_{n_w} & \mathbf{0} & \mathbf{0} & (\Phi D)^\top & \mathbf{0} & 
				\mathbf{0}   \\
				\star & \star & \gamma I_{n_x} & \mathbf{0} & \mathbf{0} & \mathbf{0} & 
				\mathbf{0} \\
				\star & \star & \star & \frac{\epsilon_1}{1 + \epsilon_1} P & \mathbf{0} & \mathbf{0} & \Phi D \Delta \\
				\star & \star & \star & \star & \frac{1}{\epsilon_1} P & \mathbf{0} & \mathbf{0}\\
				\star &  \star & \star & \star & \star & \epsilon_2 I_T & \mathbf{0} \\
				\star & \star & \star & \star & \star & \star & \frac{1}{\epsilon_2} I_{n_w}
			\end{bmatrix} \succ 0,
		\end{align}
	\end{subequations}	
	where $\star$ indicates matrix symmetry, $\Upsilon_{1,4} = ( X_1 Y + B U_0 Y)^\top$,
	$\Phi = I_{n_x} - B (N B)^{\dagger} N$, and $\epsilon_1, \epsilon_2 > 0$ are user-given scalars.
\end{theorem}
\begin{proof}
	Let $G_1 = Y P^{-1}$ and $G = [G_1, ~ G_2]$, then we obtain $Z_0 G = I_{n_z}$ from \eqref{const:control 2} and $K = U_0 G$ from \eqref{controller:thm}. 
	Applying these and $u_\text{n}(k) = K Z(x(k))$ to \eqref{eq:nominal closed sys} yields
	\begin{equation}\label{eq:lemma data closed pf1}
		x(k+1) 
		= (A Z_0 + B U_0) G Z(x(k)) + \Phi D w(k).
	\end{equation}
	Since $U_0$, $X_0$, $X_1$, $Z_0$ and $W_0$ satisfy \eqref{eq:data dynamics},  
	$X_1 = A Z_0 + D W_0$.
	Applying it to \eqref{eq:lemma data closed pf1} yields
	\begin{equation}\label{eq:lemma data closed pf2}
		x(k+1) 
		= \bar{A} x(k) + \bar{E} Q(x(k)) + \Phi D w(k),
	\end{equation}
	where $\bar{A} = (X_1  + B U_0 - D W_0) G_1$ and $\bar{E} = ( X_1 + B U_0 - D W_0) G_2$. 
	Applying \eqref{const:control 4} to \eqref{eq:lemma data closed pf2} gives
	\begin{equation}\label{eq:thm control pf4}
		x(k+1) = \bar{A} x(k) + \bar{D} \bar{w}(k)
	\end{equation}
	where $\bar{D} = \Phi D$ and $\bar{w}(k) = w(k) - W_0 G_2 Q(x(k))$.
	
	Consider the Lyapunov function $V(k) = x(k)^\top P^{-1} x(k)$. According to the Bounded Real Lemma \citep{SchererWeiland00}, \eqref{eq:thm control pf4} is $H_\infty$-robustly asymptotically 
	stable if there exists a matrix $P \succ 0$ and a scalar $\gamma > 0$ such that
	\begin{equation}\label{eq:thm control pf6}
		V(k+1) - V(k) + \gamma^{-1} \|x(k)\|^2 - \gamma \|\bar{w}(k)\|^2 < 0.
	\end{equation}
	Applying \eqref{eq:thm control pf4} to \eqref{eq:thm control pf6} and rearranging yields
	\begin{align}\label{eq:thm control pf7}	
		\setlength{\parindent}{0in}	
		&x(k)^\top \left( \bar{A}^\top P^{-1}  \bar{A} - P^{-1} + \gamma^{-1} I_{n_x} \right) x(k) 
		\nonumber\\
		& \!+\! \bar{w}(k)^\top (\bar{D}^\top P^{-1} \bar{D} - \gamma I_{n_w}) \bar{w}(k) \!+\! x(k)^\top \bar{A}^\top 
		P^{-1} \bar{D} \bar{w}(k) \nonumber\\
		&  \!+ \bar{w}(k)^\top \bar{D}^\top P^{-1} \bar{A} x(k) 
		< 0.
	\end{align}
	For any scalar $\epsilon_1 > 0$, the following inequality holds:
	\begin{align}
		& x(k)^\top \bar{A}^\top P^{-1} \bar{D} \bar{w}(k) + \bar{w}(k)^\top D^\top P^{-1} \bar{A} x(k) 
		\nonumber\\
		\leq{}& \epsilon_1^{-1} x(k)^\top \bar{A}^\top P^{-1}  \bar{A} x(k) + \epsilon_1 \bar{w}(k)^\top \bar{D}^\top P^{-1} \bar{D} \bar{w}(k). \nonumber	
	\end{align}
	Then  
	a sufficient condition for \eqref{eq:thm control pf7} is given as
	\begin{equation}\label{eq:thm control pf9}
		\xi(k)^\top	\Pi \xi(k) < 0,
	\end{equation}	
	where $\xi(k) \!=\! [x(k); \bar{w}(k)]$, $\Pi \!=\! \text{diag}(\Pi_{1,1},\Pi_{2,2})$, $\Pi_{1,1} \!=\! (1+\epsilon_1^{-1}) \bar{A}^\top P^{-1}  \bar{A} - P^{-1} + \gamma^{-1} I$ and $\Pi_{2,2} \!=\! \epsilon_1 \bar{D}^\top P^{-1} \bar{D} - \gamma I$.
	An equivalent condition to \eqref{eq:thm control pf9} is
	$-\Pi \succ 0$. Applying Schur complement \citep{SchererWeiland00} to it yields
	\begin{equation}\label{eq:thm control pf11}	
		\begin{bmatrix}
			P^{-1}  & \mathbf{0} & I_{n_x} & \bar{A}^\top & 
			\mathbf{0}\\
			\star & \gamma I_{n_w} & \mathbf{0} & \mathbf{0} & \bar{D}^\top  \\
			\star & \star & \gamma I_{n_x} & \mathbf{0} & \mathbf{0}\\
			\star & \star & \star & \frac{\epsilon_1}{1 + \epsilon_1} P & \mathbf{0} \\
			\star & \star & \star & \star & \frac{1}{\epsilon_1} P
		\end{bmatrix} 
		\succ 0.
	\end{equation}	
	Multiplying \eqref{eq:thm control pf11} with $\text{diag}(P,I,I,I,I)$ and its transpose and using $G_1 = Y P^{-1}$, we have
	%
	\begin{equation}\label{eq:thm control pf13}
		\Upsilon -  \mathcal{M} W_0^\top \mathcal{N} - \mathcal{N}^\top W_0 \mathcal{M}^\top \succ 0, 
	\end{equation}
		with 
		$\mathcal{M}^\top = [ Y, \mathbf{0}, \mathbf{0}, \mathbf{0}, \mathbf{0} ]$, 
		$\mathcal{N} = [ 
		\mathbf{0}, \mathbf{0}, \mathbf{0}, \bar{D}^\top, \mathbf{0}]$, and \\
	$
		\Upsilon =
		\begin{bmatrix}
			P  & \mathbf{0} & P & (X_1 Y + B U_0 Y)^\top  & \mathbf{0}\\
			\star & \gamma I_{n_w} & \mathbf{0} & \mathbf{0} & \bar{D}^\top  \\
			\star & \star & \gamma I_{n_x} & \mathbf{0} & \mathbf{0}\\
			\star & \star & \star & \frac{\epsilon_1}{1 + \epsilon_1} P & \mathbf{0} \\
			\star & \star & \star & \star & \frac{1}{\epsilon_1} P
		\end{bmatrix}.$\\
As shown in \cite[Lemma 4]{Persis+23}, under Assumption \ref{assume:disturbance}, $W_0 \in \mathcal{W} := \{W \in \mathbb{R}^{n_w \times T} \mid W 
W^\top \preceq \Delta \Delta^\top \}$, with $\Delta = \delta \sqrt{T} I_{n_w}$. Thus, $\mathcal{M} W^\top \mathcal{N} + \mathcal{N}^\top W \mathcal{M} \!\preceq\! \epsilon^{-1} \mathcal{M} \mathcal{M}^\top + \epsilon \mathcal{N}^\top \Delta 
	\Delta^\top \mathcal{N}$ holds for any scalar $\epsilon > 0$.  
	By this, a sufficient condition to \eqref{eq:thm control pf13} is 
	\begin{equation}\label{eq:thm control pf14}
		\Upsilon - \epsilon_2^{-1} \mathcal{M} \mathcal{M}^\top - \epsilon_2 \mathcal{N}^\top \Delta \Delta^\top \mathcal{N} \succ 0,
	\end{equation}
	for any given scalar $\epsilon_2 > 0$.		
Applying Schur complement to \eqref{eq:thm control pf14} yields \eqref{const:control 5}, which ensures \eqref{eq:thm control pf6} and guarantees the robustly asymptotic stability of \eqref{eq:thm control pf4}.
\qed
\end{proof}	

A condition for the feasibility of \eqref{op:control design} is that $Z_0$ has full row rank, seen as a requirement for data richness \citep{Persis+23}. 
The reachability and convergence of the sliding surface in Theorem \ref{theorem:sliding and convergence} depend on condition \eqref{cond:lemma sliding}, which is shown below to be satisfied by the proposed data-driven nominal control design.

\begin{proposition}\label{proposition:condition}
Under Theorem \ref{thm:control design}, \eqref{cond:lemma sliding} is satisfied with 
$\tilde{A} = N ( X_1 + B U_0 ) G_1$ and $d(k) = -W_0 G Z(x)$.
\end{proposition}
\begin{proof}
By using \eqref{eq:nominal closed sys}, \eqref{const:control 4}, \eqref{eq:lemma data closed pf2} and $u_\text{n}(k) = K Z(x(k))$, we have
$A Z(x(k)) + B K Z(x(k)) = ( X_1 + B U_0 ) G_1 x(k) - D W_0 G Z(x(k)).
$
Multiplying both its sides from the left by $N$ yields
$
 N A Z(x(k)) + N B K Z(x(k)) = \tilde{A} x(k) + N D d(k),
$
where $\tilde{A} = N ( X_1 + B U_0 ) G_1$ and $d(k) = -W_0 G Z(x)$. 
\qed
\end{proof}

The nominal controller from SDP \eqref{op:control design} builds on the data-driven nonlinearity cancellation method in \cite[Eq. (56)]{Persis+23} but differs in disturbance handling. While \cite{Persis+23} ensures robust stability of $\bar{A}$ in the system \eqref{eq:thm control pf4} with regularization to mitigate effect of $\bar{w}(k)$, we use $H_\infty$ control to enhance robustness against both $\bar{A}$ uncertainty and $\bar{w}(k)$, improving performance. Empirical studies in Section \ref{sec:simulation} also indicate better feasibility of the proposed SDP.

%% file: sections/simulation.tex

\textbf{Example 1:}
Consider an inverted pendulum system
\begin{align}\label{sim:sys1}
	x_1(k+1) ={}& x_1(k) + t_s x_2(k), \nonumber\\
	x_2(k+1) ={}& \left( 1 - \frac{t_s \mu}{m_0 \ell^2} \right) x_2(k) + \frac{t_s g}{\ell}  \sin(x_1(k)) \nonumber\\
	& + \frac{t_s}{m_0 \ell^2}  u(k) + t_s w(k), \nonumber
\end{align}
where $x_1$ is the angular displacement, $x_2$ is its velocity, $u$ is the applied torque, and $w(k)$ is a disturbance uniformly distributed in $[-\delta, \delta]$. The system parameters are sampling time $t_s = 0.1~\mathrm{s}$, mass $m_0 = 1$, length $\ell = 1$, gravity $g = 9.8$, and friction coefficient $\mu = 0.01$.


We collect $T = 30$ data samples by applying a uniformly distributed input in $[-0.5, 0.5]$. 
The proposed data-driven SMC use parameters $N = [1, 1]$, $\epsilon_1 = \epsilon_2 = 1$, $q = 0.1$,  $\sigma = 0.1$, and $\rho_1 = 0.5$. To highlight the advantages of the proposed method, we re-implement the approximate nonlinearity cancellation-based data-driven method from \cite{Persis+23} (SDP Eq. (56)), referred to as \textit{data-ANC}, for comparison.

\begin{figure}[t]
	\centering
	\includegraphics[width=0.9\columnwidth]{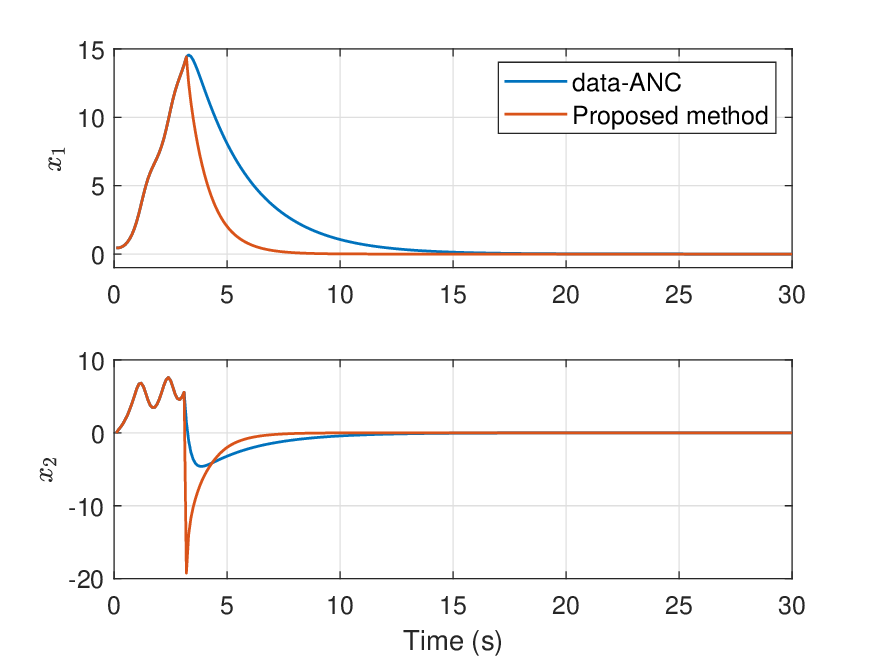} 
	\vspace{-3.5mm} 
	\caption{Performance comparison for $\delta = 0.01$: Example 1.}
	\label{fig1}
\end{figure}

\begin{figure}[t]
	\centering
	\includegraphics[width=0.9\columnwidth]{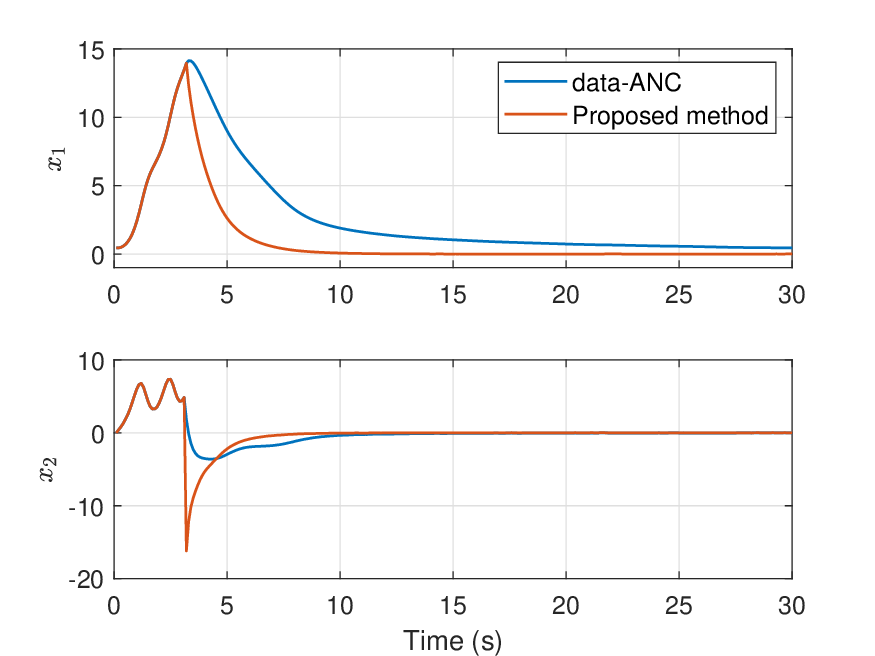}  
	\vspace{-3.5mm}
	\caption{Performance comparison for $\delta =  0.1$: Example 1.}
	\label{fig2}
\end{figure}

The performance of the proposed and \textit{data-ANC} methods is compared under varying disturbance levels (indicated by the value of $\delta$). 
The \textit{data-ANC} method is feasible up to $\delta = 0.1$, consistent with \cite{Persis+23}, while the proposed method remains feasible up to $\delta \approx 0.3$. 
Figures \ref{fig1} and \ref{fig2} illustrate results for $\delta = 0.01$ and $\delta =  0.1$, showing that the proposed method stabilizes the system faster. Notably, \textit{data-ANC} fails to drive the pendulum to the origin at $\delta = 0.1$,  whereas the proposed approach succeeds even at $\delta = 0.3$.

\begin{figure}[t]
	\centering
	\includegraphics[width=0.9\columnwidth]{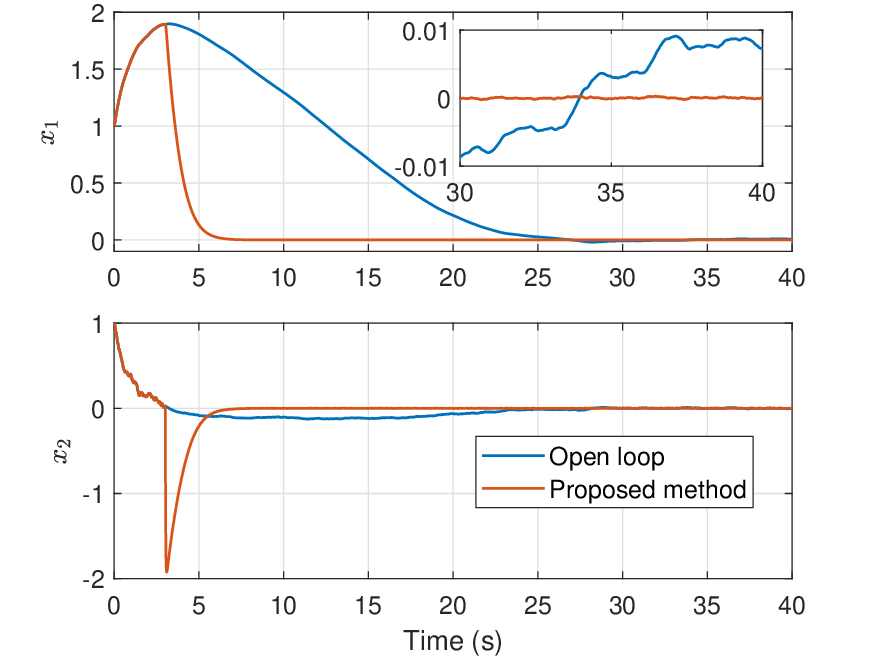}  
	\vspace{-3.5mm}
	\caption{Performance comparison: Example 2.}
	\label{fig3}
\end{figure}

\textbf{Example 2:}
Consider a cart-spring system
\begin{align}
x_1(k+1) ={}& x_1(k) + t_s x_2(k), \nonumber\\
x_2(k+1) ={}& x_2(k) - \frac{t_s k_e}{m_0} \mathrm{e}^{- x_1(k)} x_1(k) - \frac{t_s d_f}{m_0} x_2(k) \nonumber\\
&+ \frac{t_s}{m_0} u(k) + t_s w(k), \nonumber
\end{align}
where $x_1$ is the carriage displacement, $x_2$ is its velocity, $u_t$ is the external force, and $w(k)$ is a disturbance uniformly distributed in $[-\delta, \delta]$. 
The parameters are sampling time $t_s = 0.02~\text{s}$, mass $m_0 = 1$,
spring elasticity $k_e = 0.33$, and damping factor $d_f = 1$.


We collect $T = 150$ data samples by applying a uniformly distributed input in  $[-1, 1]$. The proposed control method use parameters same as Example 1. The proposed SDP \eqref{op:control design} is feasible up to the disturbance level $\delta = 0.2$, while the SDP of the \textit{data-ANC} method in \cite[Eq. (56)]{Persis+23} is infeasible even at $\delta = 0.01$. 
We compare the proposed method with the \textit{open loop} setting ($u(k) = 0$) at $\delta = 0.1$. 
As shown in Fig. \ref{fig3}, the proposed method stabilizes the system, whereas the uncontrolled system remains unstable at the origin.

%% file: sections/conclusion.tex
This paper presents a data-driven SMC for stabilizing multi-input, multi-output nonlinear systems with partially unknown dynamics and external disturbance. The design uses approximate nonlinearity cancellation, with both nominal and robust controllers being data-dependent. Simulation results demonstrate superior system stabilization and greater robustness to disturbances than the existing approximate nonlinearity cancellation-based data-driven control. Future work will explore data-driven SMC for systems with noisy data.

%% file: Data-Driven_SMC.bbl
\begin{thebibliography}{18}
\expandafter\ifx\csname natexlab\endcsname\relax\def\natexlab#1{#1}\fi
\providecommand{\url}[1]{\texttt{#1}}
\providecommand{\href}[2]{#2}
\providecommand{\path}[1]{#1}
\providecommand{\DOIprefix}{doi:}
\providecommand{\ArXivprefix}{arXiv:}
\providecommand{\URLprefix}{URL: }
\providecommand{\Pubmedprefix}{pmid:}
\providecommand{\doi}[1]{\href{http://dx.doi.org/#1}{\path{#1}}}
\providecommand{\Pubmed}[1]{\href{pmid:#1}{\path{#1}}}
\providecommand{\bibinfo}[2]{#2}
\ifx\xfnm\relax \def\xfnm[#1]{\unskip,\space#1}\fi
\bibitem[{Astolfi(2021)}]{Astolfi21n}
\bibinfo{author}{Astolfi, A.} (\bibinfo{year}{2021}).
\newblock \bibinfo{title}{Nonlinear adaptive control}.
\newblock In {\it \bibinfo{booktitle}{Encyclopedia of Systems and Control}\/}
  (pp. \bibinfo{pages}{1467--1472}).
\newblock \bibinfo{publisher}{Springer}.
\bibitem[{Berberich \& Allg{\"o}wer(2024)}]{berberich2024overview}
\bibinfo{author}{Berberich, J.}, \& \bibinfo{author}{Allg{\"o}wer, F.}
  (\bibinfo{year}{2024}).
\newblock \bibinfo{title}{An overview of systems-theoretic guarantees in
  data-driven model predictive control}.
\newblock {\it \bibinfo{journal}{Annual Review of Control, Robotics, and
  Autonomous Systems}\/},  {\it \bibinfo{volume}{8}\/}.
\bibitem[{Campi \& Savaresi(2006)}]{CampiSavaresi06}
\bibinfo{author}{Campi, M.~C.}, \& \bibinfo{author}{Savaresi, S.~M.}
  (\bibinfo{year}{2006}).
\newblock \bibinfo{title}{{Direct nonlinear control design: The virtual
  reference feedback tuning (VRFT) approach}}.
\newblock {\it \bibinfo{journal}{IEEE Transactions on Automatic Control}\/},
  {\it \bibinfo{volume}{51}\/}, \bibinfo{pages}{14--27}.
\bibitem[{Corradini(2021)}]{Corradini21}
\bibinfo{author}{Corradini, M.~L.} (\bibinfo{year}{2021}).
\newblock \bibinfo{title}{A robust sliding-mode based data-driven model-free
  adaptive controller}.
\newblock {\it \bibinfo{journal}{IEEE Control Syst. Lett.}\/},  {\it
  \bibinfo{volume}{6}\/}, \bibinfo{pages}{421--427}.
\bibitem[{De~Persis et~al.(2023)De~Persis, Rotulo \& Tesi}]{Persis+23}
\bibinfo{author}{De~Persis, C.}, \bibinfo{author}{Rotulo, M.}, \&
  \bibinfo{author}{Tesi, P.} (\bibinfo{year}{2023}).
\newblock \bibinfo{title}{Learning controllers from data via approximate
  nonlinearity cancellation}.
\newblock {\it \bibinfo{journal}{IEEE Transactions on Automatic Control}\/},
  {\it \bibinfo{volume}{68}\/}, \bibinfo{pages}{6082--6097}.
\bibitem[{De~Persis \& Tesi(2020)}]{PersisTesi20}
\bibinfo{author}{De~Persis, C.}, \& \bibinfo{author}{Tesi, P.}
  (\bibinfo{year}{2020}).
\newblock \bibinfo{title}{Formulas for data-driven control: Stabilization,
  optimality, and robustness}.
\newblock {\it \bibinfo{journal}{IEEE Transactions on Automatic Control}\/},
  {\it \bibinfo{volume}{65}\/}, \bibinfo{pages}{909--924}.
\bibitem[{De~Persis \& Tesi(2023)}]{PersisTesi23}
\bibinfo{author}{De~Persis, C.}, \& \bibinfo{author}{Tesi, P.}
  (\bibinfo{year}{2023}).
\newblock \bibinfo{title}{Learning controllers for nonlinear systems from
  data}.
\newblock {\it \bibinfo{journal}{Annual Reviews in Control}\/},  (p.
  \bibinfo{pages}{100915}).
\bibitem[{Ebrahimi et~al.(2020)Ebrahimi, Ozgoli \& Ramezani}]{ebrahimi+20}
\bibinfo{author}{Ebrahimi, N.}, \bibinfo{author}{Ozgoli, S.}, \&
  \bibinfo{author}{Ramezani, A.} (\bibinfo{year}{2020}).
\newblock \bibinfo{title}{Data-driven sliding mode control: a new approach
  based on optimization}.
\newblock {\it \bibinfo{journal}{International Journal of Control}\/},  {\it
  \bibinfo{volume}{93}\/}, \bibinfo{pages}{1980--1988}.
\bibitem[{Guo et~al.(2022)Guo, De~Persis \& Tesi}]{Guo+22d}
\bibinfo{author}{Guo, M.}, \bibinfo{author}{De~Persis, C.}, \&
  \bibinfo{author}{Tesi, P.} (\bibinfo{year}{2022}).
\newblock \bibinfo{title}{Data-driven stabilization of nonlinear polynomial
  systems with noisy data}.
\newblock {\it \bibinfo{journal}{IEEE Transactions on Automatic Control}\/},
  {\it \bibinfo{volume}{67}\/}, \bibinfo{pages}{4210--4217}.
\bibitem[{Guo et~al.(2023)Guo, De~Persis \& Tesi}]{Guo+23}
\bibinfo{author}{Guo, M.}, \bibinfo{author}{De~Persis, C.}, \&
  \bibinfo{author}{Tesi, P.} (\bibinfo{year}{2023}).
\newblock \bibinfo{title}{Data-driven control of input-affine systems via
  approximate nonlinearity cancellation}.
\newblock {\it \bibinfo{journal}{IFAC-PapersOnLine}\/},  {\it
  \bibinfo{volume}{56}\/}, \bibinfo{pages}{1357--1362}.
\bibitem[{Le{\'s}niewski(2018)}]{lesniewski2018discrete}
\bibinfo{author}{Le{\'s}niewski, P.} (\bibinfo{year}{2018}).
\newblock \bibinfo{title}{Discrete time reaching law based sliding mode
  control: A survey}.
\newblock In {\it \bibinfo{booktitle}{Proc. 22nd International Conference on
  System Theory, Control and Computing}\/} (pp. \bibinfo{pages}{734--739}).
\newblock \bibinfo{organization}{IEEE}.
\bibitem[{Lewis \& Vrabie(2009)}]{LewisVrabie09}
\bibinfo{author}{Lewis, F.~L.}, \& \bibinfo{author}{Vrabie, D.}
  (\bibinfo{year}{2009}).
\newblock \bibinfo{title}{Reinforcement learning and adaptive dynamic
  programming for feedback control}.
\newblock {\it \bibinfo{journal}{IEEE Circuits and Systems Magazine}\/},  {\it
  \bibinfo{volume}{9}\/}, \bibinfo{pages}{32--50}.
\bibitem[{Martin \& Allg{\"o}wer(2023)}]{Martin+23d}
\bibinfo{author}{Martin, T.}, \& \bibinfo{author}{Allg{\"o}wer, F.}
  (\bibinfo{year}{2023}).
\newblock \bibinfo{title}{Data-driven system analysis of nonlinear systems
  using polynomial approximation}.
\newblock {\it \bibinfo{journal}{IEEE Transactions on Automatic Control}\/},
  {\it \bibinfo{volume}{69}\/}, \bibinfo{pages}{4261--4274}.
\bibitem[{Martin et~al.(2023)Martin, Sch{\"o}n \& Allg{\"o}wer}]{Martin+23}
\bibinfo{author}{Martin, T.}, \bibinfo{author}{Sch{\"o}n, T.~B.}, \&
  \bibinfo{author}{Allg{\"o}wer, F.} (\bibinfo{year}{2023}).
\newblock \bibinfo{title}{Guarantees for data-driven control of nonlinear
  systems using semidefinite programming: A survey}.
\newblock {\it \bibinfo{journal}{Annual Reviews in Control}\/},  (p.
  \bibinfo{pages}{100911}).
\bibitem[{Riva et~al.(2024)Riva, Incremona, Formentin \&
  Ferrara}]{riva2024data}
\bibinfo{author}{Riva, G.}, \bibinfo{author}{Incremona, G.~P.},
  \bibinfo{author}{Formentin, S.}, \& \bibinfo{author}{Ferrara, A.}
  (\bibinfo{year}{2024}).
\newblock \bibinfo{title}{A data-driven approach for integral sliding mode
  control design}.
\newblock In {\it \bibinfo{booktitle}{Proc. IEEE 63rd Conference on Decision
  and Control}\/} (pp. \bibinfo{pages}{6825--6830}).
\newblock \bibinfo{organization}{IEEE}.
\bibitem[{Sarpturk et~al.(1987)Sarpturk, Istefanopulos \& Kaynak}]{Sarpturk+87}
\bibinfo{author}{Sarpturk, S.}, \bibinfo{author}{Istefanopulos, Y.}, \&
  \bibinfo{author}{Kaynak, O.} (\bibinfo{year}{1987}).
\newblock \bibinfo{title}{On the stability of discrete-time sliding mode
  control systems}.
\newblock {\it \bibinfo{journal}{IEEE Transactions on Automatic Control}\/},
  {\it \bibinfo{volume}{32}\/}, \bibinfo{pages}{930--932}.
\bibitem[{Scherer \& Weiland(2000)}]{SchererWeiland00}
\bibinfo{author}{Scherer, C.}, \& \bibinfo{author}{Weiland, S.}
  (\bibinfo{year}{2000}).
\newblock \bibinfo{title}{Linear matrix inequalities in control}.
\newblock {\it \bibinfo{journal}{Lecture Notes, Dutch Institute for Systems and
  Control, Delft, The Netherlands}\/},  {\it \bibinfo{volume}{3}\/}.
\bibitem[{Tang \& Daoutidis(2022)}]{TangDaoutidis22}
\bibinfo{author}{Tang, W.}, \& \bibinfo{author}{Daoutidis, P.}
  (\bibinfo{year}{2022}).
\newblock \bibinfo{title}{Data-driven control: Overview and perspectives}.
\newblock In {\it \bibinfo{booktitle}{Proc. American Control Conference}\/}
  (pp. \bibinfo{pages}{1048--1064}).
\newblock \bibinfo{organization}{IEEE}.

\end{thebibliography}
